\documentclass[10pt]{article}
\usepackage{graphicx}

\def\Title#1{\begin{center} {\Large #1 } \end{center}}
\def\Author#1{\begin{center}{ \sc #1} \end{center}}
\def\Address#1{\begin{center}{ \it #1} \end{center}}

\newcommand\pubblock{\rightline{\begin{tabular}{l} Proceedings of the Fifth Annual LHCP\\ \pubnumber\\
         \pubdate  \end{tabular}}}

\newenvironment{Abstract}{\begin{quotation} \begin{center} 
             \large ABSTRACT \end{center}\bigskip 
      \begin{center}\begin{large}}{\end{large}\end{center} \end{quotation}}

\newenvironment{Presented}{\begin{quotation} \begin{center} 
             PRESENTED AT\end{center}\bigskip 
      \begin{center}\begin{large}}{\end{large}\end{center} \end{quotation}}

\def\beq{\begin{equation}}
\def\eeq#1{\label{#1}\end{equation}}
\def\eeqn{\end{equation}}

\def\beqa{\begin{eqnarray}}
\def\eeqa#1{\label{#1}\end{eqnarray}}
\def\eeqan{\end{eqnarray}}

\let\bar=\overbar

\def\Dslash{\not{\hbox{\kern-4pt $D$}}}
\def\dslash{\not{\hbox{\kern-2pt $\del$}}}

\def\msb{{\bar{\ssstyle M \kern -1pt S}}}

\usepackage{color}
\usepackage{subfig}
\usepackage{lineno}

\newcommand\pubnumber{ ATL-PHYS-PROC-2017-093 }

\newcommand\pubdate{\today}

\def\affiliation{
On behalf of the ATLAS Collaboration, \\
LAPP, CNRS/IN2P3, ENIGMASS Labex and Université Savoie Mont Blanc, Annecy-le-Vieux, France}

\begin{document}
\large
\begin{titlepage}
\pubblock

\vfill
\Title{ Di- and multiboson measurements in ATLAS }
\vfill

\Author{ Elena Yatsenko  }
\Address{\affiliation}
\vfill
\begin{Abstract}

Measurements of the di- and multiboson  production cross sections at the LHC constitute stringent tests of the electroweak sector of the Standard Model and provide a model-independent means to search for new physics at the TeV scale.
The ATLAS collaboration has performed studies of $ZZ$, $WZ$, $WW$, $WW\gamma$ and $WZ\gamma$ productions in various decay modes at $\sqrt{s}=8$ and 13~TeV, including total, fiducial and differential cross-section measurements. These results are compared to state-of-the art theory predictions and are used to 
provide constraints on new physics, by setting limits on anomalous gauge boson couplings.

\end{Abstract}
\vfill

\begin{Presented}
The Fifth Annual Conference\\
 on Large Hadron Collider Physics \\
Shanghai Jiao Tong University, Shanghai, China\\ 
May 15-20, 2017
\end{Presented}
\vfill
\end{titlepage}
\def\thefootnote{\fnsymbol{footnote}}
\setcounter{footnote}{0}
%

\normalsize 


\section{Introduction}

Studies of processes containing multiple electroweak bosons ($W$, $Z$ or $\gamma$) in  final states offer a powerful probe of electroweak (EWK) sector of the Standard Model (SM) and provide an important test of quantum chromodynamics (QCD) theory calculations.  
In many extensions of the SM the new phenomena involve the presence of new resonances, that couple to bosons. Even if the scale of the new physics is beyond the reach of the LHC, the multiboson production could still be affected in the form of a modification of the couplings strength of gauge boson self-interactions, giving rise to anomalous triple and quartic gauge boson couplings (aTGC and aQGC).
Moreover, the SM di- and multiboson processes constitute an important background for direct searches of new physics processes that are forming multiboson final states. 
Since diboson processes can proceed via a Higgs boson propagator, non-Higgs-mediated diboson production represents an important background in studies of the Higgs boson properties. The presented diboson and multiboson measurements are based on the proton-proton collision data collected by the ATLAS detector~\cite{Aad:2008zzm} at the LHC at centre-of-mass energies of $\sqrt{s}=8$ and 13~TeV.

\section{Di- and multiboson production cross section measurements}

\begin{figure}[!b]
\centering
	\subfloat[]{\hspace{-3.5mm}\includegraphics[height=0.3\textwidth]{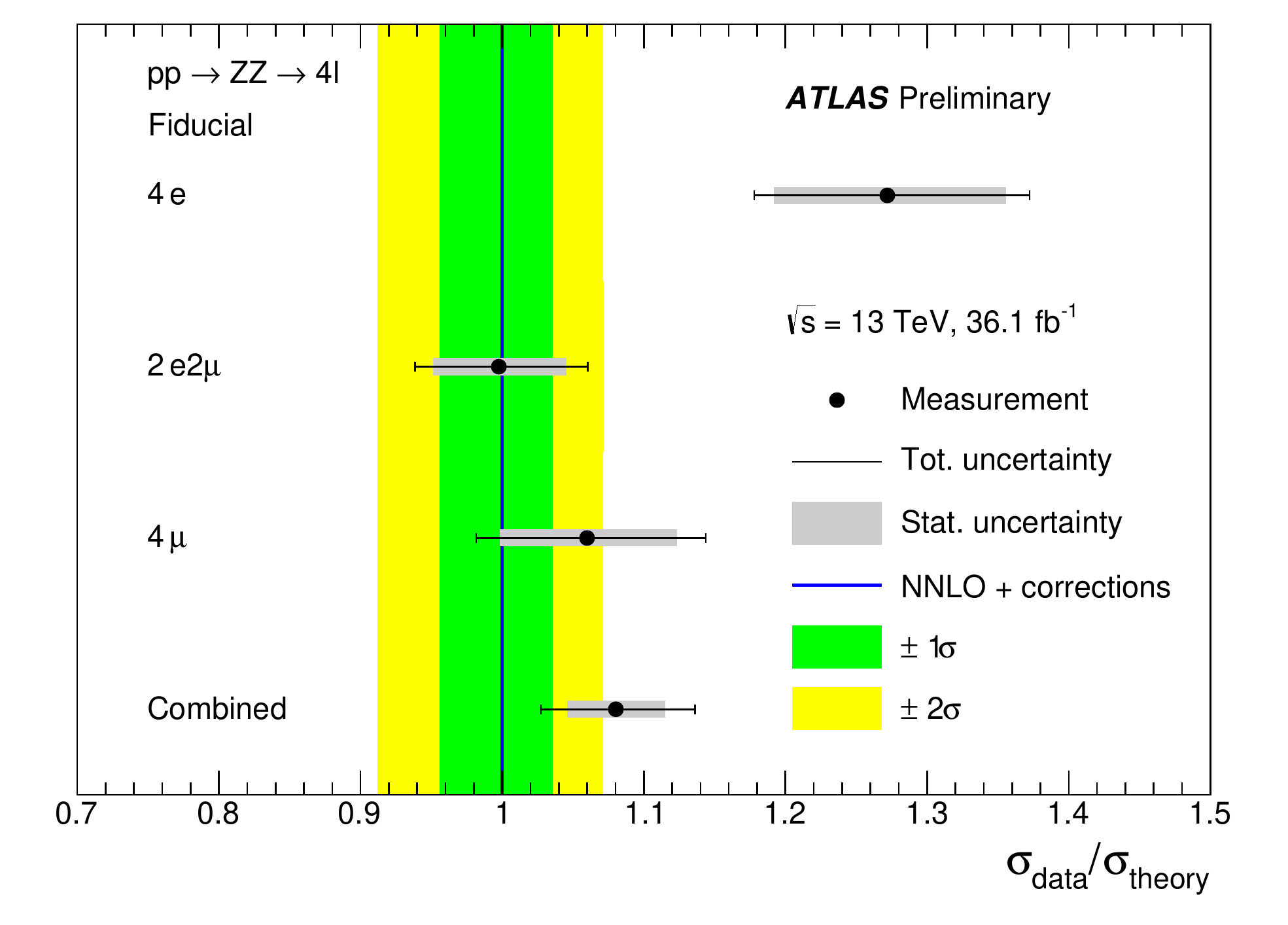}}
	\subfloat[]{\includegraphics[height=0.31\textwidth]{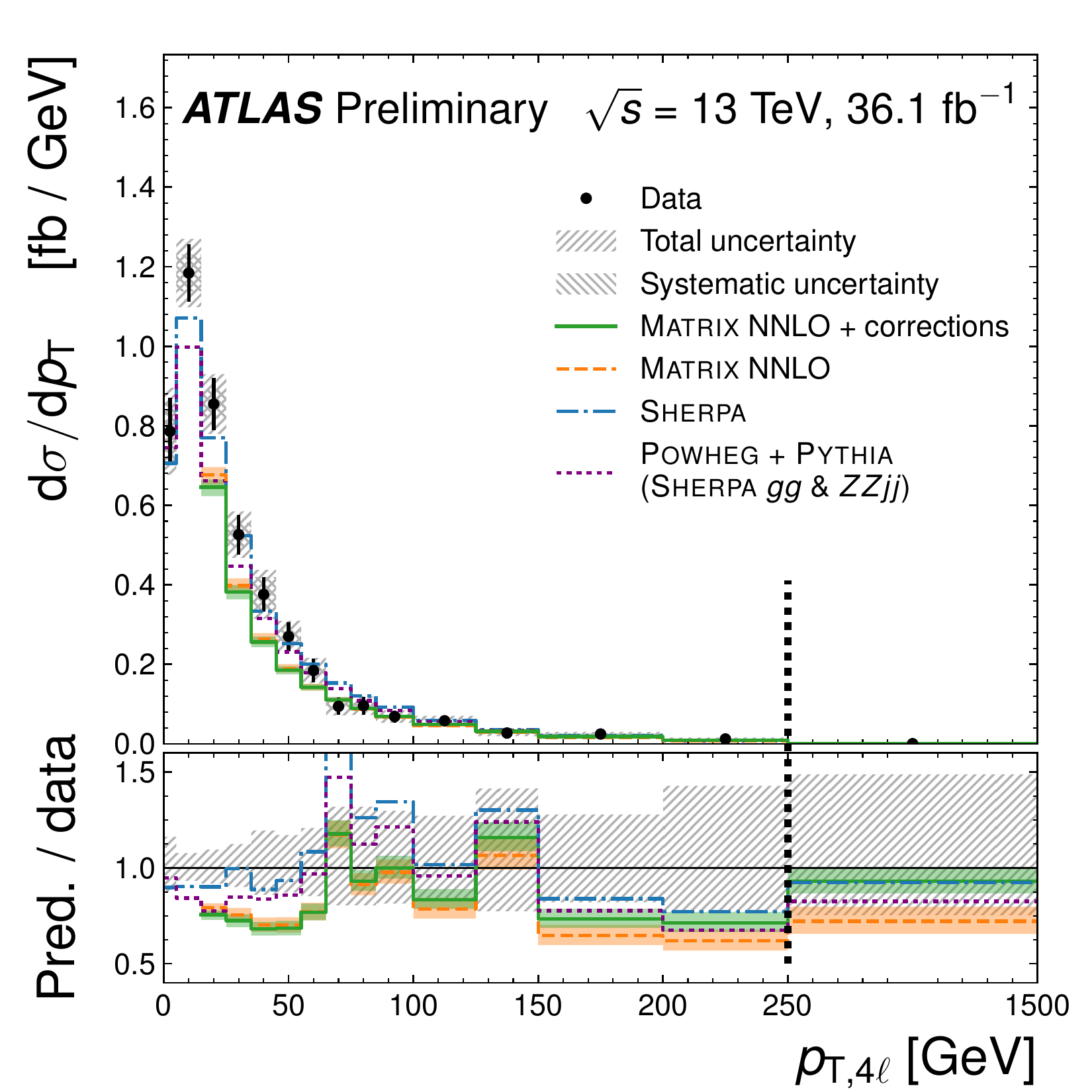}}
	\subfloat[]{\includegraphics[height=0.31\textwidth]{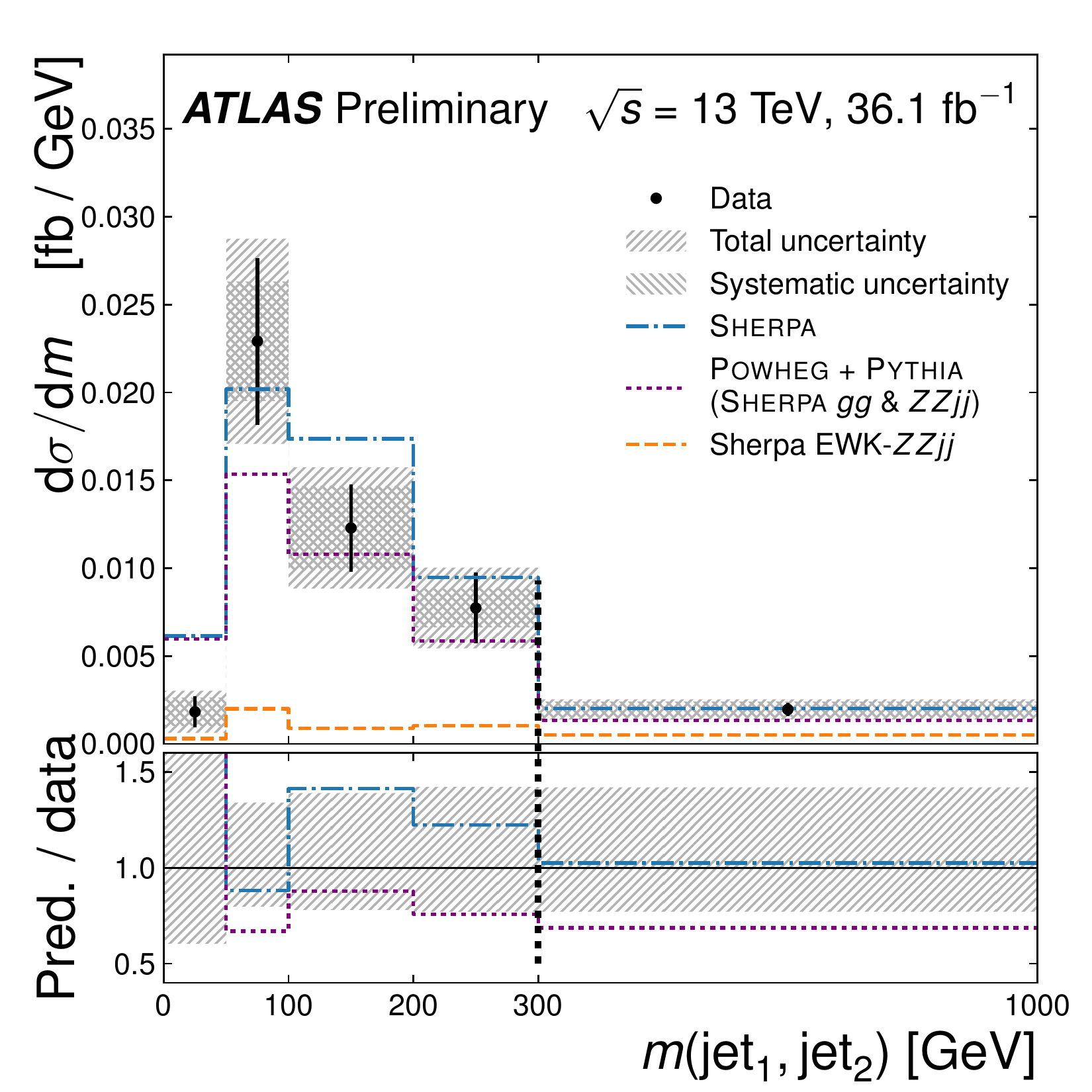}}
\caption{ Comparison of the measured $ZZ$ fiducial cross sections to SM predictions: (a) integrated cross section, and differential cross section (b) for the transverse momentum of the four-lepton system,  and (c) for the invariant mass of the two leading-$p_{\tt{T}}$ jets~\cite{ATLAS:2017eyk}. }
\label{fig:ZZ_CS}
\end{figure}

The $ZZ$ boson pair production cross section has been measured by ATLAS using 36.1~fb$^{-1}$ of data collected during 2015 and 2016~\cite{ATLAS:2017eyk}. The measurement is performed in the fully leptonic decay mode, with $Z$ bosons decaying into electrons or muons. On-shell $Z$ bosons considered in this measurement imply a superposition of a $Z$ boson and a virtual photon in the mass range between 66 and 116~GeV. The experimental signature of four isolated high-$p_{\tt{T}}$ leptons 
allows to select a very clean signal sample with background contamination of approximately 2\% of the total signal yield.
The cross section in the fiducial phase space is measured with 5\% accuracy, including the statistical uncertainty of 3.2\%, and as shown in Figure~\ref{fig:ZZ_CS}(a) is in a good agreement with the SM calculations at NNLO QCD. With a signal sample of 1017 observed events, the kinematics of $ZZ$ events is studied in details by measuring the production cross section differentially as a function of twenty different observables, including measurements of additional jet activity in the event. In Figures~\ref{fig:ZZ_CS} (b) and (c) are shown examples of the differential cross section results - the transverse momentum distribution of the four-lepton system, $p_{\tt{T},4\ell}$, which provides information about QCD and EWK radiation by measuring recoil against all other particles produced in the collision, and the invariant mass of the two leading-$p_{\tt{T}}$ jets, $m(\textrm{jet}_1,\textrm{jet}_2)$, which is particularly sensitive to the rare $ZZ$ weak boson-pair production through vector-boson scattering, that has not been experimentally observed so far.

\begin{figure}[t]
\centering
	\subfloat[]{\includegraphics[width=0.44\textwidth]{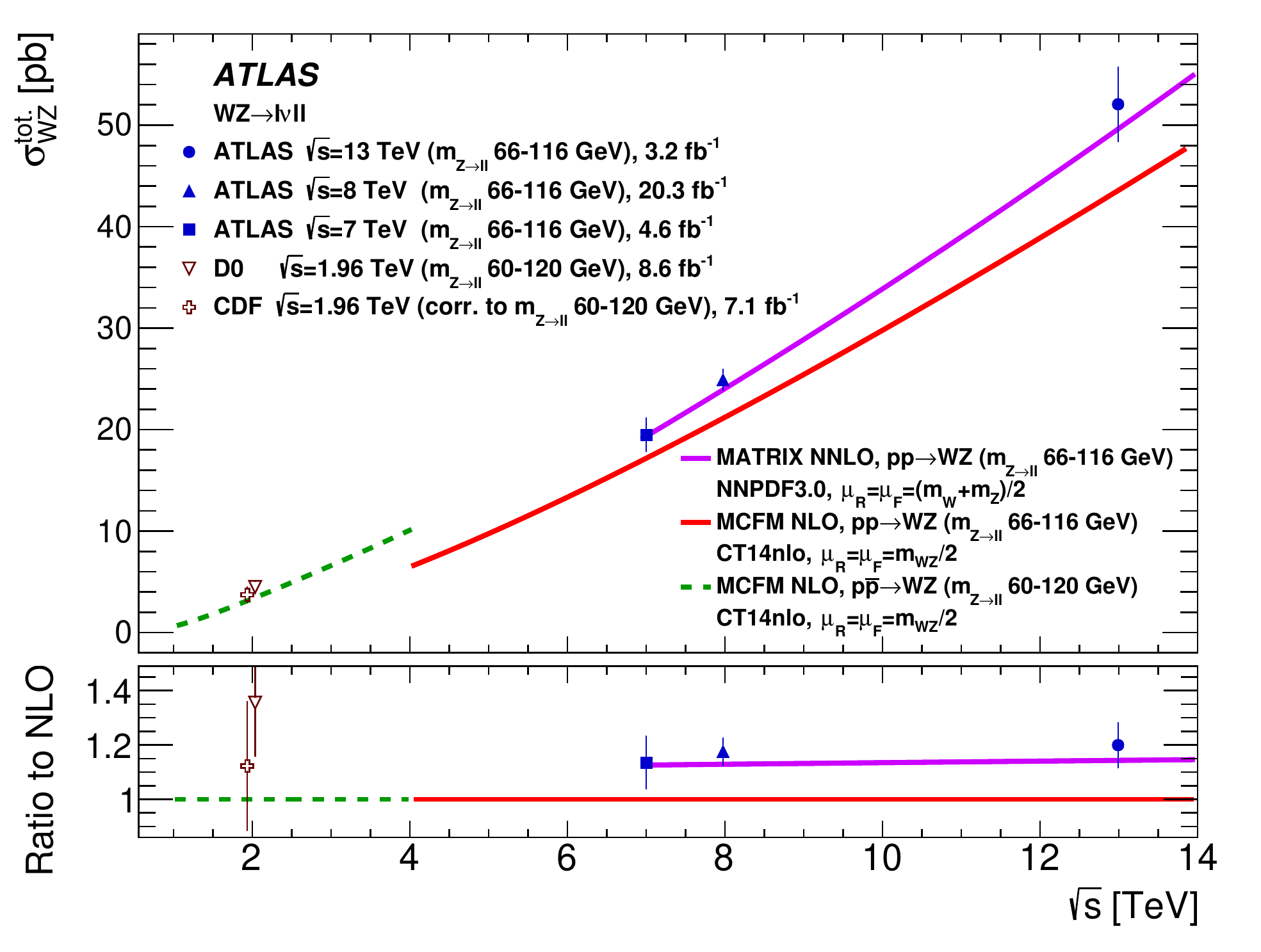}}
	\subfloat[]{\includegraphics[width=0.29\textwidth]{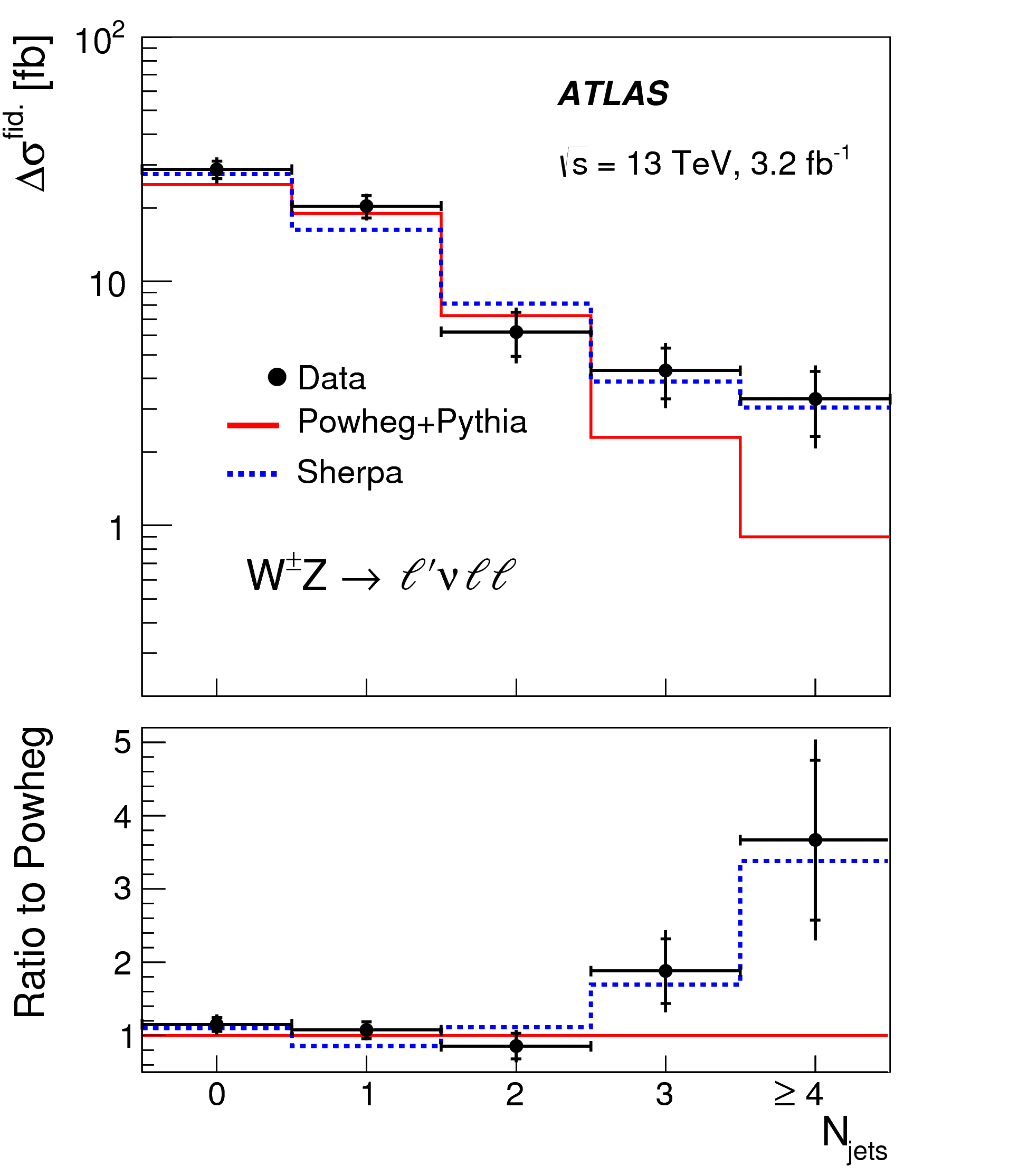}}
	\subfloat[]{\includegraphics[width=0.29\textwidth]{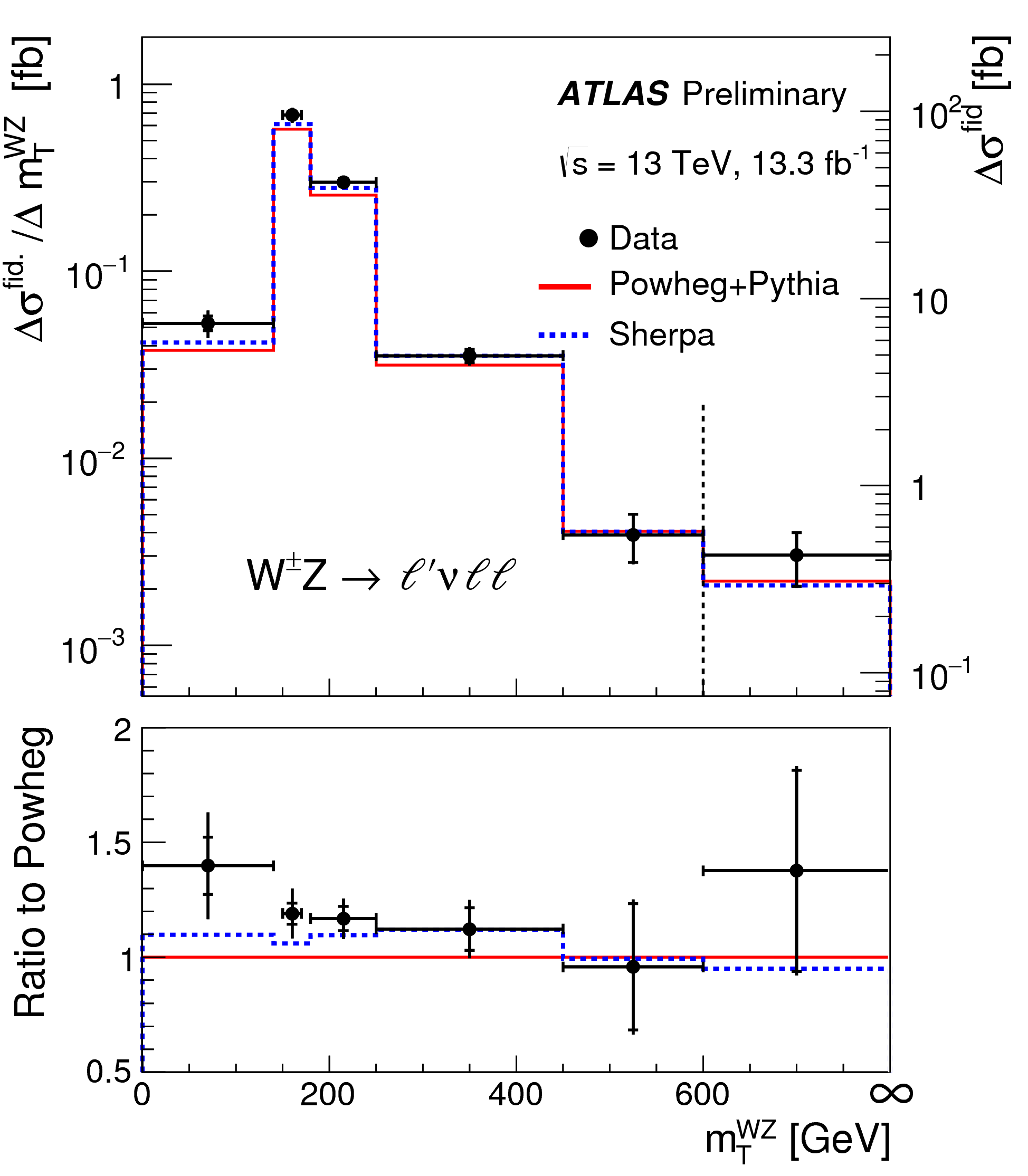}}
\caption{ Comparison of $W^{\pm}Z$ cross section measurements with SM theory predictions: (a) in the total phase space and in the fiducial phase space measured differentially as a function of (b) exclusive jet multiplicity and (c) of the transverse mass variable $m_{\tt{T}}^{WZ}$ for the $WZ$ system~\cite{Aaboud:2016yus,ATLAS:2016qzn}. }
\label{fig:WZ_CS}
\end{figure}

The fully leptonic $W^{\pm}Z$ production cross section has been measured with 3.2 and 13.3~fb$^{-1}$ of $\sqrt{s}=13$~TeV data~\cite{Aaboud:2016yus,ATLAS:2016qzn}. The measurements are performed in the $\ell'\nu\ell\ell$ final states, where the gauge bosons decay to electrons or muons. With the integrated luminosity of 13.3~fb$^{-1}$ the signal sample of 2417 observed events is consistent with the expectation of $2315\pm11$ events and is contaminated by about 21\% of background contribution. The background dominantly comes from events with non-prompt leptons and jets misidentified as leptons. The fiducial cross section is measured with a 7\% accuracy, the main experimental uncertainties being associated to the estimation of the background from fake leptons, and the lepton reconstruction and identification efficiencies. 
The comparison of the measured cross section extrapolated to the total phase space with the SM theory prediction is shown in Figure~\ref{fig:WZ_CS}(a). 
The roughly 20\% underestimate of the data by the NLO QCD predictions is resolved with the NNLO QCD calculations, that provide a very good description of the data at different centre-of-mass energies.
Examples of the differential cross section measurement for the $W^{\pm}Z$ production are shown in Figures~\ref{fig:WZ_CS}(b) and (c). The cross section as a function of exclusive jet multiplicity provides an important test of the perturbative QCD for diboson processes, and the differential cross section as a function of the transverse mass of the $WZ$ system, $m_T^{WZ}$, is particularly interesting due to its sensitivity to possible new physics effects. 
 
\begin{figure}[htb]
\centering
\includegraphics[width=0.41\textwidth]{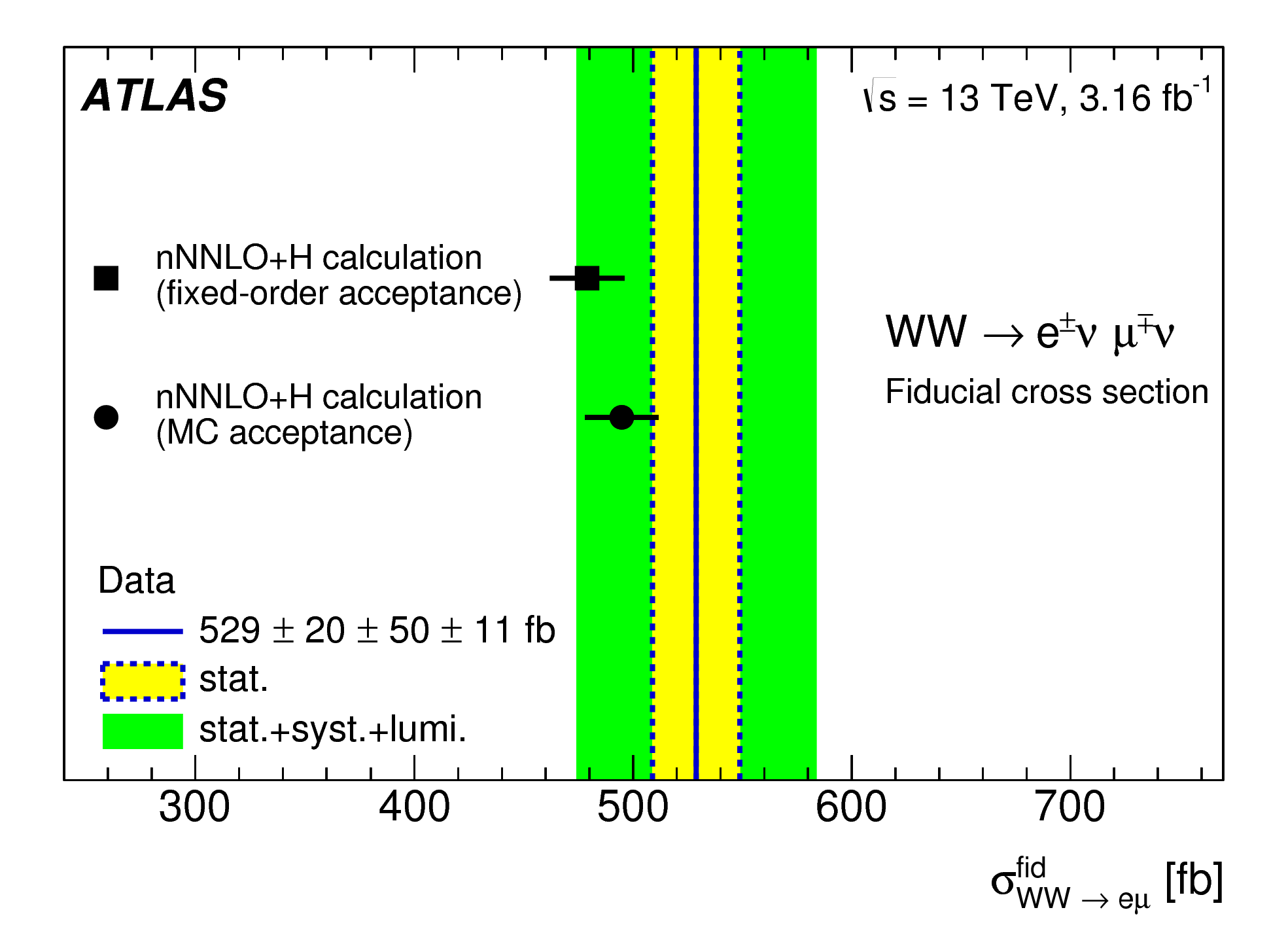}
\includegraphics[width=0.41\textwidth]{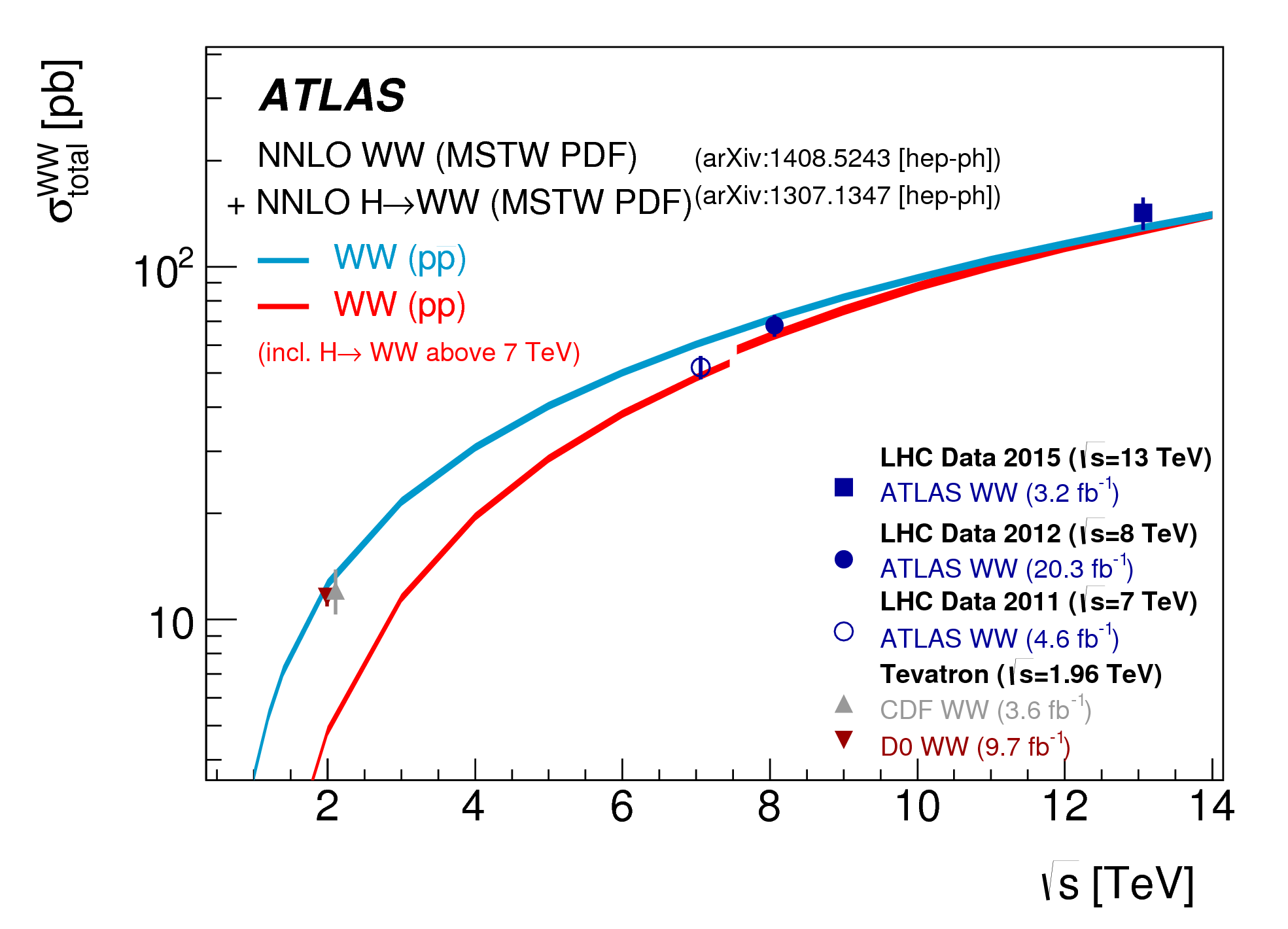}
\caption{ Comparison of the $W^+W^-$ production cross section with the theoretical predictions in (a) fiducial phase space and (b) in the total phase space~\cite{Aaboud:2017qkn}.}
\label{fig:WW_CS}
\end{figure}

The production of opposite-charge $W$-boson pairs decaying into an electron, a muon and neutrinos, $W^{\pm}W^{\mp}\rightarrow e^{\pm}\nu\mu^{\mp}\nu$, is measured using 3.16~fb$^{-1}$ of $\sqrt{s}=13$~TeV data~\cite{Aaboud:2017qkn}. The requirement of two final state leptons with different flavors allows to suppress the Drell-Yan background contamination, while the background from top-quark processes is reduced by discarding events with reconstructed high-$p_{\tt{T}}$ jets. The resonant $gg\rightarrow H\rightarrow WW$ process is included as a part of the signal. The signal selection results in 1351 observed $W^{\pm}W^{\mp}$-candidate events and is consistent with $1351\pm37$ expected events with about 26\% background contamination, where the dominant background is due to $t\bar{t}$ and single top processes. The total uncertainty on the measured fiducial cross section is 11\% with the main uncertainty coming from the jet selection and jets energy scale and resolution. The comparison of the measured fiducial and total cross sections with state-of-the art theoretical calculations is shown in Figure~\ref{fig:WW_CS}.

\begin{figure}[b]
\centering
\includegraphics[width=0.41\textwidth]{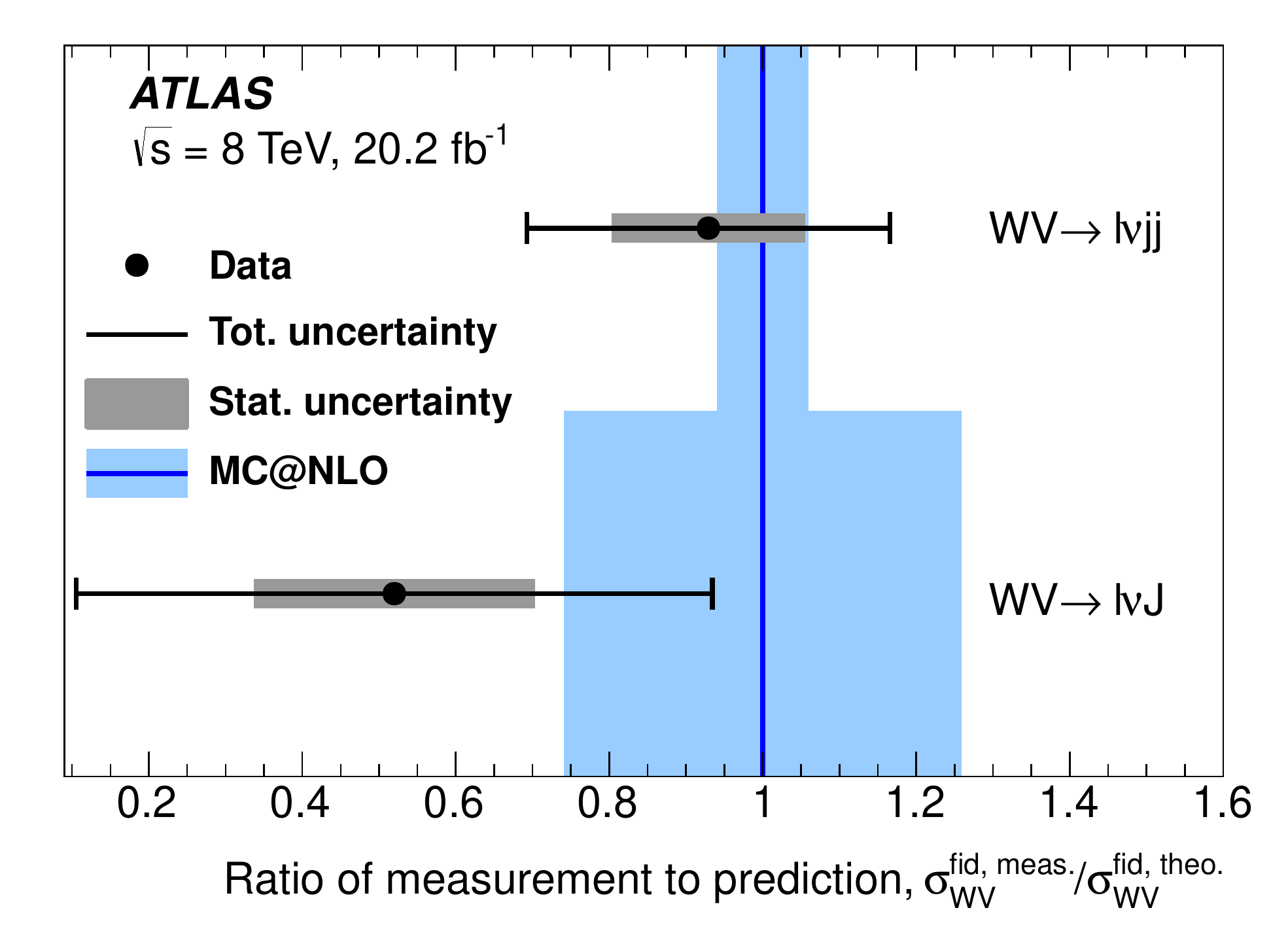}
\caption{ The ratios of the measured $WV$ fiducial cross-sections in the semileptonic final states to the SM NLO QCD predictions~\cite{Aaboud:2017cgf}.}
\label{fig:WV_CS}
\end{figure}

Measurement of $WW/WZ\rightarrow\ell\nu qq'$ boson pair production, with one $W$ boson decaying leptonically to $e\nu$ or $\mu\nu$ and one $W$ or $Z$ boson decaying hadronically, has been performed using 20.2~fb$^{-1}$ of $\sqrt{s}=8$~ TeV data~\cite{Aaboud:2017cgf}. The diboson production in the semileptonic channel has about six times higher branching fractions than the fully leptonic channels. However, the experimental precision in this measurement is limited by the large background contamination from $W+$jets and $t\bar{t}$ processes
with the signal-to-background ratio of 5-10\%. The measurements are performed for two event topologies of hadronically decaying boson - with two resolved jets reconstructed as separate objects, $WV\rightarrow\ell\nu jj$, and with two boosted jets reconstructed as a single jet of a large radius, $WV\rightarrow\ell\nu J$. 
The measured fiducial cross section in the resolved topology is $\sigma_{\textrm{fid}}(WV\rightarrow\ell\nu jj) = 209\pm28(\textrm{stat.})\pm45(\textrm{syst.})$~fb, and in the boosted topology is $\sigma_{\textrm{fid}}(WV\rightarrow\ell\nu J) = 30\pm11(\textrm{stat.})\pm22(\textrm{syst.})$~fb. The comparison of the measured cross sections with the theoretical predictions at NLO QCD is shown in Figure~\ref{fig:WV_CS}, both agree with the theory within uncertainties. Since in some events two jets can be reconstructed both as separate objects and as a single large-radius jet, the two cross section measurements are performed in partially overlapping phase spaces and thus no combination of the two measurements is performed.

$WW\gamma$ and $WZ\gamma$ triboson production has been studied using 20.2~fb$^{-1}$ of $\sqrt{s}=8$~TeV data~\cite{Aaboud:2017tcq}. Fully leptonic and semileptonic final states are studied with one $W$ decaying leptonically and one $W$ or $Z$ boson decaying hadronically or leptonically. The $WW\gamma$ production cross-section is measured in a fully leptonic final state containing a muon, an electron, a photon and neutrinos. 
Upper limits on the production cross section are derived for both  fully leptonic and  semileptonic final states. The signal selection in the fully leptonic $WW\gamma$ measurement leads to 26 observed events with the signal-to-background ratio of 81\%.
The dominant background to the fully leptonic triboson process comes from $t\bar{t}\gamma$ production. The semileptonic channel is highly contaminated by $W\gamma$+jets background. The signal selection yields 490 observed events in electron and 599 observed events in muon channels with signal-to-background ratios of 2.5\% and 2.9\%, respectively.
The measured fiducial cross section in the $WW\gamma\rightarrow e\nu\mu\nu\gamma$ channel of $\sigma_{\textrm{fid}}=1.5\pm0.9(\textrm{stat.})\pm0.5(\textrm{syst.})$~fb is in agreement with the NLO QCD theory prediction of 2.0~fb. 
The dominant experimental uncertainty is due to the jet energy scale.
The observed (expected) significance of this cross section corresponds to 1.4~$\sigma$ (1.6~$\sigma$).

\section{Search for anomalous gauge couplings}

Triple and quartic self-interactions of the EWK gauge bosons are completely fixed in the SM by the gauge structure of the EWK theory and any deviation from the SM values would indicate presence of new physics beyond the SM.
An effective field theory approach~\cite{Degrande:2012wf} is intensively used to parametrise constraints on possible new physics effects along with an effective Lagrangian~\cite{Hagiwara:1986vm} framework for charged aTGCs and an effective  vertex  function  approach~\cite{Baur:2000cx} for neutral gauge couplings. 
These frameworks provide model-independent ways of searching for physics beyond the SM. EWK boson couplings deviating from gauge constraints lead to an enhancement of the production cross section at high energy scales of the interactions thus affecting the tails of the differential distributions. No deviations from the SM predictions are observed in the above described measurements  and confidence intervals on aTGC and aQGC parameters are derived.

\begin{figure}[t]
\centering
\includegraphics[width=0.4\textwidth]{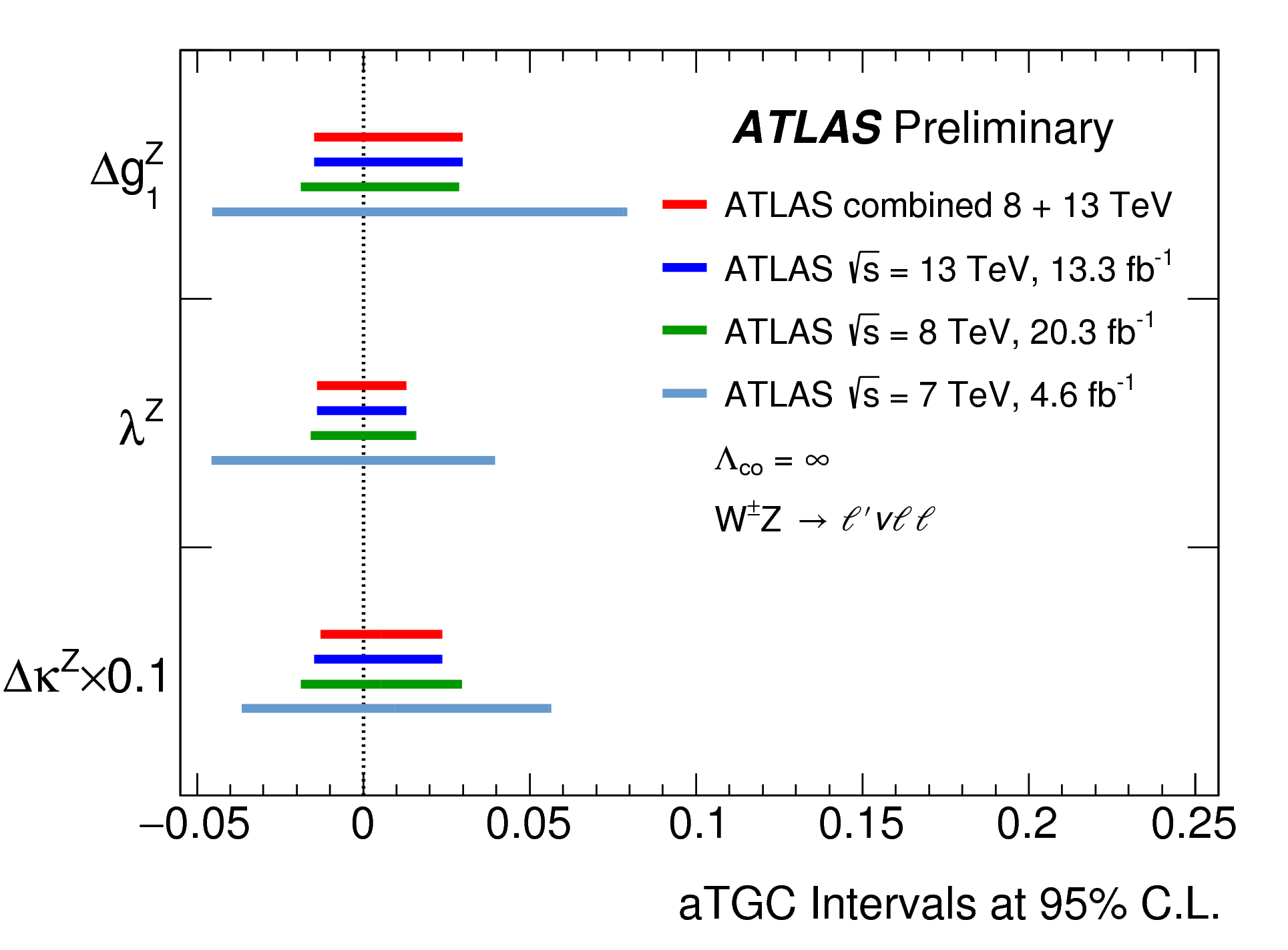}
\caption{ The observed confidence intervals at 95\% CL on charged aTGCs derived from fully leptonic $W^{\pm}Z$ measurements~\cite{ATLAS:2016qzn}.}
\label{fig:WZ_aTGC}
\end{figure}

\begin{figure}[]
\centering
\includegraphics[width=0.3\textwidth]{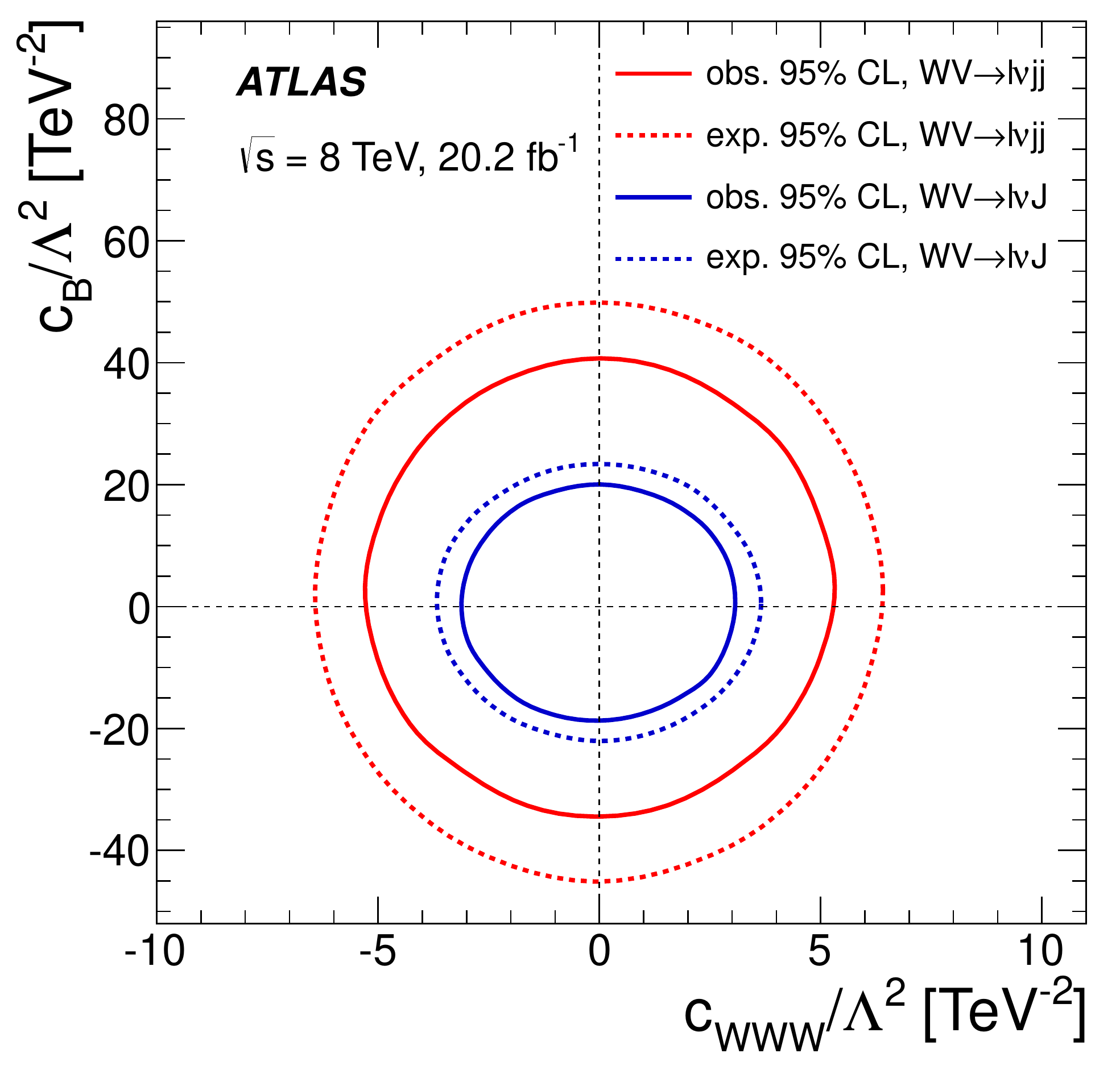}
\includegraphics[width=0.3\textwidth]{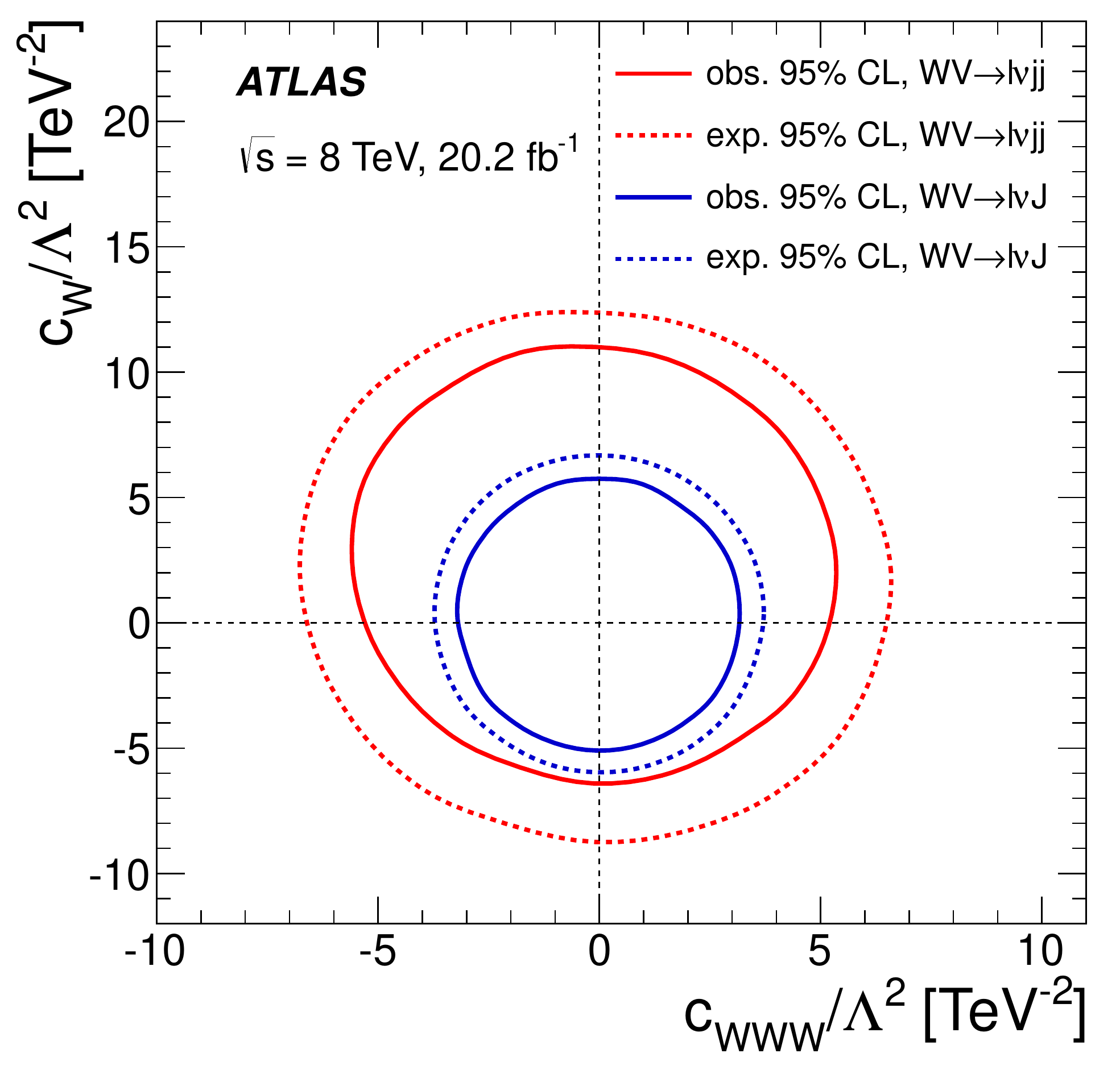}
\includegraphics[width=0.3\textwidth]{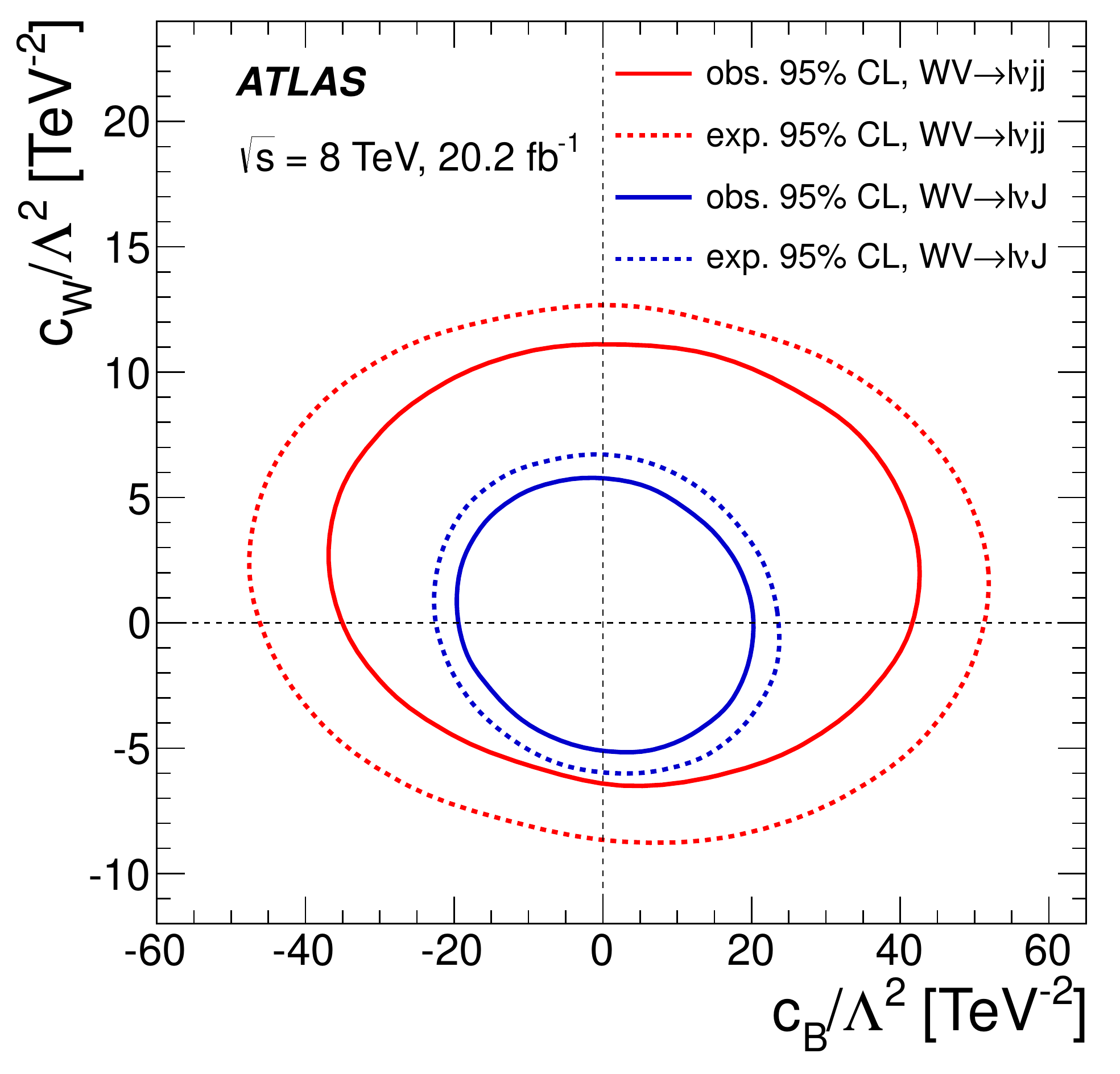}
\caption{ The observed and expected confidence intervals at 95\% CL on charged aTGCs derived from the semileptonic $WW$/$WZ$ resolved (red lines) and boosted (blue lines) analyses~\cite{Aaboud:2017cgf}. }
\label{fig:WV_aTGC}
\end{figure}

\begin{figure}[h]
\centering
\begin{minipage}[b]{0.45\linewidth}
  \includegraphics[width=\textwidth]{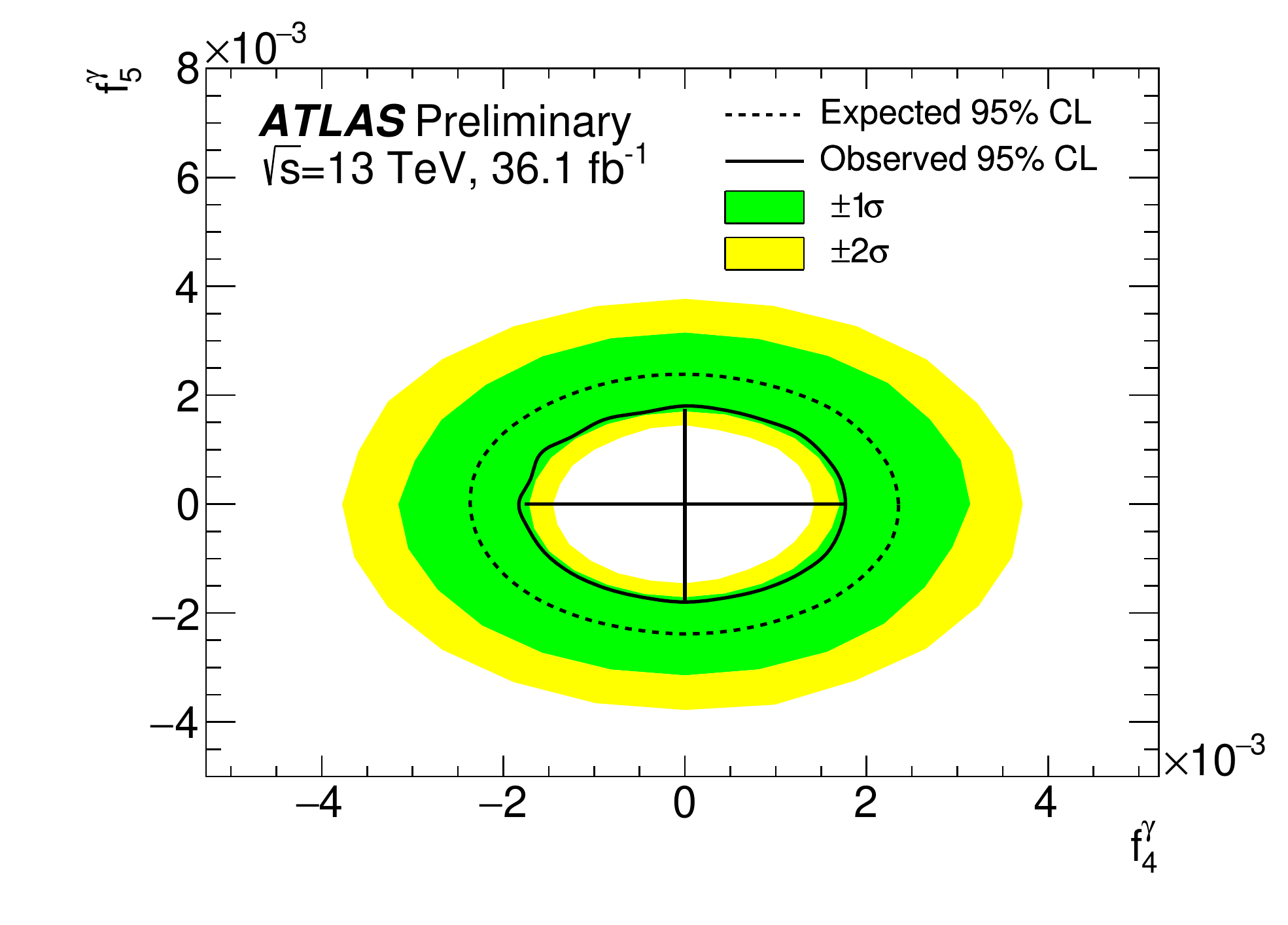}
  \caption{An example of the observed and expected confidence intervals at 95\% CL on neutral aTGCs derived from the $ZZ$ measurement~\cite{ATLAS:2017eyk}.}\label{fig:ZZ_limit}
\end{minipage}  
\quad
\begin{minipage}[b]{0.42\linewidth}
  \includegraphics[width=\textwidth]{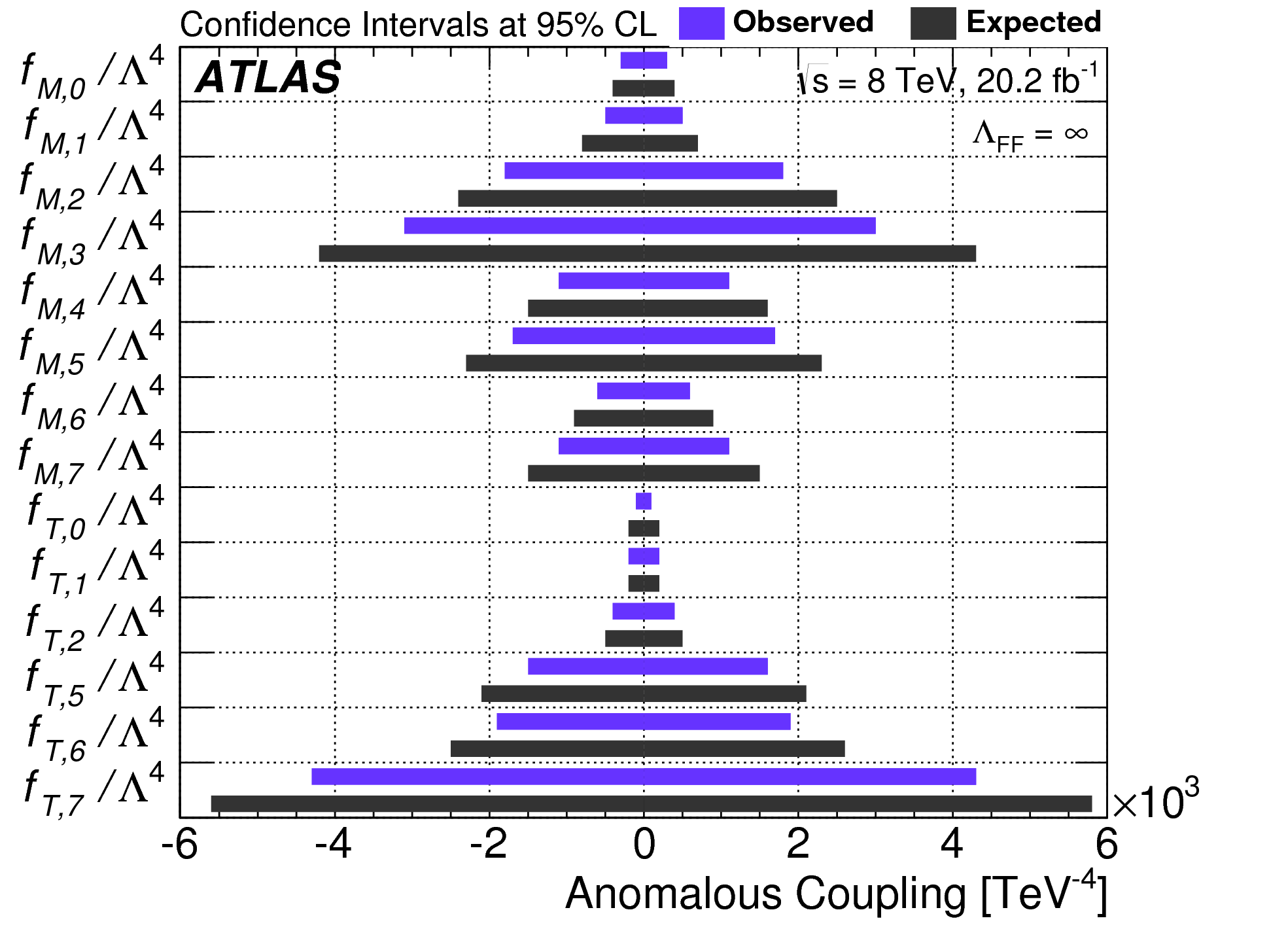}
  \caption{The observed and expected confidence intervals at 95\% CL on aQGCs derived from the $WW\gamma$ and $WZ\gamma$ measurements~\cite{Aaboud:2017tcq}.}\label{fig:aQGC}
\end{minipage}  
\end{figure}

Searches of charged aTGCs for $WWZ$ and $WW\gamma$ vertices are performed using $WZ$ and $WW$ measurements. The two processes provide complementary sensitivity to the charged aTGC parameters, since $WZ$ production is sensitive only to $WWZ$ couplings while $WW$ can also be produced involving a photon and thus probing $WW\gamma$ couplings.   
In the fully leptonic $W^{\pm}Z$ analysis the constraints are derived from the reconstructed transverse mass spectrum of the $WZ$ system. The combination of 13~TeV $W^{\pm}Z$ measurements with 8~TeV results allows to improve previously derived constraints by factors of up to 1.2 as shown in Figure~\ref{fig:WZ_aTGC}.
Using the semileptonic $WW$ and $WZ$ measurements the constraints are derived from the $p_{\tt{T}}$ spectrum of the dijet system in the resolved analysis, and from the $p_{\tt{T}}$ distribution of the large-radius jet in the boosted analysis. The boosted topology of the semileptonic measurement probes higher boson $p_{\tt{T}}$ thus allowing to access larger energy scale of the interaction and providing higher sensitivity to aTGCs as shown in Figures~\ref{fig:WV_aTGC}.
A search for neutral aTGCs, $ZZZ$ and $ZZ\gamma$, which are forbidden at tree-level in the SM, are performed using the $ZZ$ cross section measurement. The constraints are derived from the reconstructed transverse momentum distribution of the leading-$p_{\tt{T}}$ $Z$ boson candidate, which is one of the observables with the highest sensitivity to predicted neutral aTGC effects. 
No indication of neutral aTGCs were found and an example of the two-dimentional confidence intervals is shown in Figure~\ref{fig:ZZ_limit}.
Searches for aQGCs are performed using the $WW\gamma$ and $WZ\gamma$ measurement in the high-$E_{\tt{T}}$ photon region. The results are found to be consistent with the SM prediction of zero as shown in Figure~\ref{fig:aQGC}. 

\section{Conclusions}

Using the data collected at centre-of-mass energies of $\sqrt{s}=8$ and 13~TeV various measurements of diboson and multiboson production have been performed by the ATLAS collaboration. The results include a number of unfolded differential cross sections.
The data are found to be consistent with the SM predictions, and no indication of anomalous couplings is observed. The measurements are used to derive confidence intervals on aTGC and aQGC parameters.

\end{document}